\documentclass[aps,showpacs,twocolumn,unsortedaddress,superscriptaddress]{revtex4-1}
\usepackage{graphics, bm, psfrag, amsmath, amssymb, epsfig, grffile, float}
\usepackage{color}
\usepackage{hyperref}
\hypersetup{
    colorlinks=true, %set true if you want colored links
    linktoc=all,     %set to all if you want both sections and subsections linked
    linkcolor=blue,  %choose some color if you want links to stand out
}

\usepackage{booktabs}

\begin{document}

\title{Computational search for Dirac and Weyl nodes in $f$-electon antiperovskites}
\author{Anna Pertsova}
\affiliation{Nordita, KTH Royal Institute of Technology and Stockholm University, Roslagstullsbacken 23, SE-106 91 Stockholm, Sweden}
\author{R. Matthias Geilhufe}
\affiliation{Nordita, KTH Royal Institute of Technology and Stockholm University, Roslagstullsbacken 23, SE-106 91 Stockholm, Sweden}
\author{Martin Bremholm}
\affiliation{Department of Chemistry and iNANO, Aarhus University, 8000 Aarhus, Denmark}
\affiliation{Center for Materials Crystallography, Aarhus University, 8000 Aarhus, Denmark}
\author{Alexander V. Balatsky}
%\affiliation{Institute for Materials Science, Los Alamos National Laboratory, Los Alamos New Mexico 87545, USA}
\affiliation{Nordita, KTH Royal Institute of Technology and Stockholm University, Roslagstullsbacken 23, SE-106 91 Stockholm, Sweden}
\affiliation{Dept. of Physics, University of Connecticut,  Storrs, CT 06269, USA}

\date{\today}

%-------------------------------------------------------

\begin{abstract}
We present the result of an \textit{ab initio} search for new 
Dirac materials among inverse perovskites. Our investigation is focused on the less studied class of lanthanide antiperovskites 
 containing heavy $f$-electron elements in the cation position. Some of the studied compounds have not yet been synthesized experimentally. Our computational 
approach is based on density functional theory calculations which account for spin-orbit interaction and strong correlations of the $f$-electron atoms. 
We find several promising candidates among lanthanide antiperovskites 
which host bulk Dirac states close to the Fermi level. 
Specifically, our calculations reveal massive three-dimensional Dirac states in materials of the class A$_3$BO, where A=Sm, Eu, Gd, Yb and B=Sn, Pb. 
In materials with finite magnetic moment, such as Eu$_3$BO (B=Sn, Pb), the degeneracy of the Dirac nodes 
is lifted, leading to appearance of Weyl nodes.

\end{abstract}

%------------------------------------------------------
\maketitle

\section{Introduction}
Dirac materials (DMs) is a growing class of materials which exhibit a linear, Dirac-like spectrum of quasiparticle 
excitations~\cite{wehling2014dirac}. Examples of DMs that have attracted particular attention in the past decade include graphene
~\cite{neto2009electronic}, three-dimensional (3D) topological insulators (TIs)~\cite{hasan_rmp2010,qi_rmp2011}, topological 
crystalline insulators (TCIs)~\cite{Fu_prl2011_TCI}, and the newly discovered three-dimensional DMs such as Dirac~\cite{Young_prl2012,neupane2014dirac} and 
Weyl semimetals~\cite{Wan_prb2011,huang2015weyl,xu2015weyl}. 
Over the past several years, it has been suggested that a new subclass of DMs can be found in cubic antiperovskite 
materials~\cite{Klintenberg2014,Sun2010,Kariyado_jpsj2011,Hsieh_prb2014,Chiu_prb2017}

Antiperovskites, or inverse perovskites, are inorganic compounds with a perovskite type structure in which the positions of cations and anions are interchanged. 
The typical structure is A$_3$BX, where A is an electropositive cation, B is a divalent metallic anion and X is a 
 monovalent anion i.e. the position of A and X are reversed compared to ordinary perovskites. Antiperovskites have  
 a great potential for electronic, magnetic and thermoelectric applications~\cite{Bilal_APV2015}. 
 
Regarding the search for DMs, much focus has been on antiperovskite oxides with a simple cubic structure, % (see Fig.~\ref{fig1}),  
 A$_3$BO, where A is an alkaline earth metal such as Ca, Ba, or Sr. The prototypical example is Ca$_3$PbO which 
 was proposed as a potential DM in a number of studies~\cite{Sun2010,Kariyado_jpsj2011,Hsieh_prb2014,Chiu_prb2017}. 
Recently, superconductivity has been reported in a similar compound, the antipervoskite Dirac-metal oxide Sr$_3$SnO with hole doping~\cite{Maeno_natcomm2016}. 
There has been an increasing interest in other antiperovskite materials. A group of nitride antiperovskites with a common structure A$_3$BiN, where A=Ca, Ba, Sr has 
been predicted to be 3D TIs when subject to properly designed uniaxial strain~\cite{Sun2010}. 
A recent density functional theory (DFT) based study of a larger class of alkaline earth pnictides A$_3$BX, where A=Ca, Ba, Sr and B, X=N, P, As, Sb, Bi found  
several materials that can be driven into topological phases by properly engineered strain~\cite{Goh_prb2018}. A promising candidate is Ca$_3$BiP which is a 
topological semimetal without strain but can be driven into a 3D TI or a Dirac semimetal phase.

The nature of the Dirac states in antiperovskite materials predicted so far has been studied from different viewpoints. 
Initially, the data mining search based on electronic 
structure calculations performed in Ref.~\cite{Klintenberg2014} identified Ca$_3$PbO (and similar compounds, A$_3$BO, A=Ca, Ba, Sr and B=Sn, Pb) as a possible 3D TI. 
More specifically, it 
was suggested to be a strong 3D TI based on apparent 
band inversion at the $\Gamma$ point. The band inversion was confirmed in other studies~\cite{Sun2010,Kariyado_jpsj2012,Hsieh_prb2014}. 
However, despite the band inversion, the product of parities of the bands at time-reversal invariant momenta remains the same with or without spin orbit coupling (SOC). 
Therefore, based on the parity criterion~\cite{fu2007topological_2}, 
 strain free Ca$_3$PbO is a trivial insulator. Properly engineered uniaxial strain can change the ordering of the inverted bands, turning the material into a
 3D TI, similarly to Ca$_3$BiN~\cite{Sun2010}. In the absence of strain,   
the Ca$_3$PbO family was predicted to be TCI with unusual surface states with open Fermi surface similar to Weyl semimetals~\cite{Hsieh_prb2014,Chiu_prb2017}. 
At the same time, it was suggested that 
 Ca$_3$PbO is a massive 3D DM, with Dirac nodes occurring in the 3D Brillouin away from 
any high-symmetry point~\cite{Kariyado_jpsj2011,Kariyado_jpsj2012,Kariyado_phd2012,Kariyado_jpcs2015,Fuseya_jpsj2015} . 

In fact, the topological gap at $\Gamma$ and the 3D Dirac states away from $\Gamma$  
 may coexist in these materials. 
Based on topological band theory, Hsieh \textit{et al.}~\cite{Hsieh_prb2014} showed that the band inversion at the $\Gamma$ point, 
responsible for the TCI phase in Ca$_3$PbO, leads to the appearance of a gapped node (avoided crossing) along the $\Gamma$-$X$ direction of the 3D Brillouin zone. 
 This feature was studied in detail by Kariyado and Ogata~\cite{Kariyado_jpsj2011,Kariyado_jpsj2012} using a combination 
 of \textit{ab initio} and tight-binding methods. In particular, 
 it was shown that these states can be described as massive Dirac fermions.
Photoemission experiments are required to verify these theoretical predictions. 

More recent theoretical work has predicted the existence of Dirac nodal lines in materials with weak SOC, which can be realized in 
some antiperovskite compounds~\cite{Yu_prl2015,Kim_prl2015}. 
Specifically, Yu \textit{et al.}~\cite{Yu_prl2015} found that in the absence of SOC, Cu$_2$PdN is a nodal-line semimetal with three nodal circles due to cubic symmetry. 
Nodal lines originate from the band inversion at the $R$ point of the 3D Brillouin zone and are protected by 
 time reversal and inversion symmetries~\cite{Weng_prb2015}. Inclusion of SOC drives the system into a Dirac semimetal phase with three pairs of Dirac nodes. 
 Similar conclusions were found by Kim \textit{et al.}~\cite{Kim_prl2015} for a more general case of Cu$_2$N doped with non-magnetic transition metal atoms, 
 i.e. in Cu$_3$X$_x$N (antiperovskite structure) with X=Ni, Cu, Zn, Pd, Ag, Cd. In particular, two maximally-doped cases, Cu$_3$PdN and Cu$_3$ZnN were studied 
 with effective Hamiltonian and \textit{ab initio} methods, demonstrating bulk Dirac nodal lines and nearly-flat surface states.

 We note that the existence of nodal lines is not a universal property of anitiperovskites. Since it requires vanishing or weak spin-orbit interaction, 
it is not realized in compounds containing elements with very large atomic numbers. On the other hand, band inversion at time reversal invariant momenta and the occurrence of nodes 
along adjacent symmetry lines is a common feature. Unlike in 3D TIs, the band inversion in antiperovskites is not due to SOC but can be induced by changing the lattice
constant or the chemical elements. The details of the electronic structure, such as 
the character of the inverted bands and the position of the Dirac nodes are material specific.%, for example it is different for Ca$_3$PbO and Cu$_3$PdN families.

In this work we focus on the scarcely studied group of lanthanide antiperovskite oxides, %many of which have not been synthesized to date. 
 A$_3$BO, where A=Sm, Eu, Gd, Yb and B=Sn, Pb. 
%We consider two subgroup, (i) A$_3$BO, where A=Sm, Eu, Gd, Yb and B=Sn, Pb, and (ii) A$_3$BO, where A=La, Pr, Nd, and B=In, Tl. 
Among published work, Yb$_3$SnO and Yb$_3$PbO were predicted to be 3D TIs~\cite{Klintenberg2014}. In a recent work 
based on a large-scale \textit{ab intio} search for topological materials, Yb$_3$PbO has been identified as a possible TCI.~\cite{Zhang_Nature2019}
Based on electronic structure calculations,   we find that this group of compounds is characterized by massive 
Dirac nodes in the 3D Brillouin zone. The most interesting candidates are Eu$_3$SnO, Eu$_3$PbO, Yb$_3$SnO and Yb$_3$PbO which have Dirac states near the Fermi level. 
Among these materials, Yb$_3$PbO is the most promising for further studies and applications since the Dirac states are isolated from other bands.    
The origin of the 3D Dirac states is similar to the case of 
Ca$_3$PbO. Our DFT calculations show that most of the considered lanthanide antiperovskites  
possess a finite magnetic moment due to unpaired $f$-electrons in their outer shells (Yb$_3$BO is an exception 
due to the filled \textit{f}-shell of Yb).  
This offers intriguing possibilities of combining magnetism and Dirac fermion physics in pristine materials without doping 
or proximity effects. Furthermore, in magnetic lanthanide antiperovskites, e.g. Eu$_3$BO (B=Sn, Pb), based on the number of the nodes and the band degenercy, 
we identify the nodes in the 3D Brillouin zone as massive Weyl nodes.

The paper is organized as follows. In Sec.~\ref{methods} we provide computational details. Section~\ref{results} 
 reports the outcome of our \textit{ab initio} search for DMs among lanthanide antiperovskites. Namely, we summarize the results of electronic structure calculations 
 for the materials of this class in the bulk phase. Several candidates displaying Dirac/Weyl states are identified. 
The electronic structure of the most promising candidates with isolated cones near the Fermi level are 
 discussed in more detail. We also analyze the character and the possible origin of the 
Dirac states using a specific example of Yb$_3$PbO. Finally, we draw some conclusions in Sec.~\ref{concl}.

\section{Methods}\label{methods}

\begin{figure}[ht!]
\centering
\includegraphics[width=0.8\linewidth,clip=true]{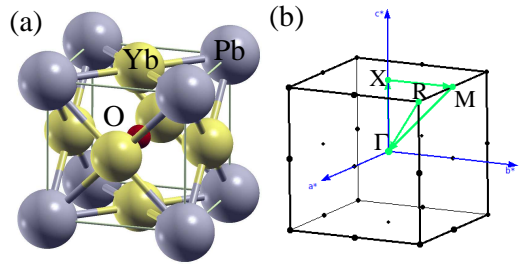}
\caption{(a) The cubic antiperovskite structure of Yb$_3$PbO. (b) R-$\Gamma$-X-M-$\Gamma$ path in the Brillouin zone used for banstructure calculations.}
\label{fig0}
\end{figure}

We performed DFT calculations using the full-potential all-electron linearized augmented plane-waves method as implemented in the  
Wien2k \textit{ab initio} package~\cite{Wien2k}. The Perdew-Burke-Ernzerhof generalized gradient approximation (PBE-GGA)~\cite{Perdew_prb1996} is used for
the exchange correlation functional. For lanthanide antiperovskites,  
the GGA+U method was used to account for local correlations at the cation site. 
The Hubbard U parameter in the GGA+U approach effectively describes the on-site repulsion associated with the narrow 3\textit{d} or 4\textit{f} bands.
We used the following arguments to choose a suitable value of U for our calculations of lanthanide antiperovskites.

In a standard GGA calculation with U=0, the 4$f$ band appears right at the Fermi level. A finite U pushes the band away from the Fermi level into the valence band. 
Since the experimental data for 4$f$ valence states is not available, it is sufficient to chose the value of U such as (i) the 4$f$  band is pushed down into the valence band, (ii) varying U 
in a reasonable range does not affect the electonic states at the Fermi level, in particular the Dirac nodes. We perfomed test calculations 
for one of the compunds (Yb$_3$PbO) with U ranging from 3eV to 12eV. We verified that changing U does not affect 
the nodes, while it shifts the narrow 4$f$ states to lower energy.  The overall bandstructure around the Fermi level also does not change appreciably.  
Furthermore, comparing total energies of the system calculated with different U, we found that U=10~eV corresponds to a local minimum of the total energy. 
We also noted that for U=8 and 9~eV, the arrangement of bands in the vicinity of the $\Gamma$ point is slightly different comapared to other cases. 
For all other values of U, the bands in the energy window $[$-2~eV,2~eV$]$ 
are essentially identical. Hence, we choose U=10~eV,  which 
is within the commonly accepted range for rare earths~\cite{Marel_Sawatzky_prb1988} and, in particular, lanthanides.~\cite{Pang_comp_mater_sci_2011, Khadraoui_opt_mater_2015, Khadraoui_chem_phys_2015}.

We used $126$ non-equivalent k-points in a $10\times{10}\times{10}$ mesh of the 
first Brillouin zone for self-consistent calculations.   SOC is included in all calculations as it can not be 
ignored in heavy-element compounds.  Geometry optimization was performed to find the equilibrium lattice constant, which is particularly important for compounds for which no experimental 
structural data is available (Sm$_3$SnO, Sm$_3$PbO, Gd$_3$SnO, Gd$_3$PbO). 
The Brillouin zone path used for banstructure calculations is shown in Fig.~\ref{fig0}(b).

The results of structural (volume) optimization together with experimental (estimated or observed) lattice constants for lanthanide antiperovskites are presented in 
Table~\ref{tab:lattice}. For compounds that have been synthesized and for which structural data can be found in the literature (Eu$_3$SnO, Eu$_3$PbO, Yb$_3$SnO, 
Yb$_3$PbO), the DFT optimized lattice constants are close to experimental values. We use experimental lattice parameters~\cite{Velden_APVs_2004,Nuss_APVs_2015} 
for the existing lanthanide antiperovskites, as well for the test data set of alkali antiperovskites~\cite{Widera_APVs_1980, Nuss_APVs_2015}. 

\begin{table}[ht]
\caption{Experimental and DFT optimized lattice constant for lanthanide antiperovskite oxides: A$_3$BO, A=Sm, Eu, Gd, Yb, and B=Sn, Pb; space group 
 $\mathrm{P\bar{m}3m}$. E(O) refers to estimated (observed) structural data. References are given for the experimental lattice constants.}
\begin{tabular}{l||c|c|c|c|c|r}
System &  Experiment, a ($\AA$) & DFT, $a$ ($\AA$) \\
\hline
Sm$_3$SnO   & 5.111 (E) & 4.984 \\
\hline 
Sm$_3$PbO   & 5.124 (E) & 5.015  \\
\hline
\hline
Eu$_3$SmO   & 5.077 (O)~\cite{Nuss_APVs_2015} &  5.018 \\
\hline 
Eu$_3$PbO   & 5.091 (O)~\cite{Nuss_APVs_2015} & 5.051 \\
\hline
\hline
Gd$_3$SnO   & 5.043 (E)  & 4.851  \\
\hline 
Ga$_3$PbO   & 5.058 (E) & 4.887   \\
\hline 
\hline
Yb$_3$SnO   & 4.837 (O)~\cite{Velden_APVs_2004} & 4.804  \\
\hline 
Yb$_3$PbO   & 4.859 (O)~\cite{Velden_APVs_2004}  & 4.842  \\
\end{tabular}
\label{tab:lattice}
\end{table}

\section{Results and Discussion}\label{results}

We use the well studied Ca$_3$PbO group of antiperovskite oxides containing alkali metals as a test set for our \textit{ab initio} search. 
The results are summarized in Table~\ref{tab:alkali} [see also the electronic structure for Ca$_3$PbO in Fig.~\ref{fig2}(a,b)]. 
 The table includes information 
on the presence (+) or absence (-) of Dirac nodes according to our DFT electronic structure calculations, as well as the availability of 
 structural data (synthesis), and
references to previous theoretical work with a brief note on the predicted character of the Dirac states. 

Our results for the Ca$_3$PbO family are in agreement with Kariyado and Ogata~\cite{Kariyado_jpsj2011}.  
Namely, we find gapped nodes formed by linearly dispersing bands along the $\Gamma$-$X$ direction in all compounds of this group. The nodes are located 
at the Fermi level. In Ba$_3$SnO and Ba$_3$PbO, there are other states crossing the Fermi level thus obscuring the Dirac states.
 The results also largely agree with Hsieh \textit{et al.}~\cite{Hsieh_prb2014}, with the possible exception of 
 Ca$_3$SnO. 

\begin{table}[ht]
\caption{Possible Dirac materials (DMs) in antiperovskite oxides with alkali metals: A$_3$BO, A=Ca, Ba, Sr, B=Sn, Pb; space group 
$\mathrm{P\bar{m}3m}$.}
\begin{tabular}{l||c|c|c|c|r}
System &  DM & Synthesis & Theory \\
\hline
\hline
 Ba$_3$PbO & + & \cite{Widera_APVs_1980,Velden_APVs_2004,Nuss_APVs_2015} & 
\begin{tabular}{@{}c@{}}  \cite{Klintenberg2014}(3DTI), \cite{Sun2010}(3DTI+strain), \\ 
\cite{Kariyado_jpsj2011}(3DDM), \cite{Hsieh_prb2014}(TCI)
\end{tabular}\\
\hline
 Ba$_3$SnO & -  & \cite{Widera_APVs_1980,Velden_APVs_2004,Nuss_APVs_2015}  &  \cite{Klintenberg2014,Sun2010,Kariyado_jpsj2011,Hsieh_prb2014} \\
\hline
\hline
 Ca$_3$PbO  & +  & \cite{Widera_APVs_1980,Velden_APVs_2004,Nuss_APVs_2015}  & \begin{tabular}{@{}c@{}} \cite{Klintenberg2014}(3DTI), \cite{Sun2010}(3DTI+strain), \\ 
\cite{Kariyado_jpsj2011}(3DDM), \onlinecite{Hsieh_prb2014}(TCI)
\end{tabular}  \\
\hline 
Ca$_3$SnO   &  +   & \cite{Widera_APVs_1980,Velden_APVs_2004,Nuss_APVs_2015}  &  \cite{Klintenberg2014}(3DTI), \cite{Kariyado_jpsj2011}(3DDM), \cite{Hsieh_prb2014}(TCI) \\
\hline
\hline
Sr$_3$PbO   & +   & \cite{Widera_APVs_1980,Velden_APVs_2004,Nuss_APVs_2015}  &  \begin{tabular}{@{}c@{}} \cite{Klintenberg2014}(3DTI), \cite{Sun2010}(3DTI+strain), \\ 
\cite{Kariyado_jpsj2011}(3DDM), \cite{Hsieh_prb2014}(TCI)
\end{tabular}  \\
\hline 
Sr$_3$SnO   & +  & \cite{Widera_APVs_1980,Velden_APVs_2004,Nuss_APVs_2015}  & \cite{Klintenberg2014}(3DTI), \cite{Hsieh_prb2014}(TCI)  \\
\end{tabular}
\label{tab:alkali}
\end{table}

We now focus on lanthanide antiperovskite oxides. The results of the electronic structure calculations for this group of materials are summarized 
in Table~\ref{tab:lanthanides}, where we also include the information on the value of the calculated magnetic moment (for lanthanide atoms in the unit cell). 
The available experimental and theoretical literature 
is considerably more scarce compared to the Ca$_3$PbO family. Structural data is available for Eu$_3$BO and Yb$_3$BO (B=Sn, Pb), while for 
Sm$_3$BO and Gd$_3$BO (B=Sn, Pb), we use the lattice constant obtained from DFT (see Table~\ref{tab:lattice}). 
%We note that Yb$_3$BO (B=Sn, Pb) was classified as a potential TI by Klintenberg \textit{et al.}~\cite{Klintenberg2014}.
%, based on the inversion of band character of valence 
%and conduction bands at the $\Gamma$ point, similarly to Ca$_3$PbO.  

\begin{table}[ht]
\caption{Possible Dirac materials (DMs) in lanthanide antiperovskite oxides: A$_3$BO, A=Sm, Eu, Gd, Yb, B=Sn, Pb; space group 
 $\mathrm{P\bar{m}3m}$.}
\begin{tabular}{l||c|c|c|c|c|r}
System &  DM & \begin{tabular}{@{}c@{}} Magnetic \\ moment ($\mu_\mathrm{B}$)\end{tabular}   & Synthesis & Theory \\
\hline
\hline
Sm$_3$PbO   & - & 5.8   &  - \\
\hline 
Sm$_3$SnO   & - &  5.7 & - &  - \\
\hline
\hline
Eu$_3$PbO   & + & 6.9 & \cite{Velden_APVs_2004}& - \\ % Schnyder et al. (unpubl.)  \\
\hline 
Eu$_3$SnO   & + & 6.9  & \cite{Velden_APVs_2004} & - \\
\hline
\hline
Gd$_3$PbO   & -  & 7.1  & - &  - \\
\hline 
Ga$_3$SnO   & - & 7.2  & - & -  \\
\hline 
\hline
Yb$_3$PbO   & +  & 0 & \cite{Nuss_APVs_2015} & \cite{Klintenberg2014}(3DTI) \\
\hline 
Yb$_3$SnO   & +   & 0 & \cite{Nuss_APVs_2015} &  \cite{Klintenberg2014}(3DTI) \\
%\hline
%\hline
%La$_3$InO  & ? & ? & No & Yes &  No \\
%\hline
%La$_3$TlO   & ? & ? & No & \textbf{No} &  No  \\
%\hline 
%Nd$_3$InO   & No & Yes  & No & \textbf{No} &  No \\
%\hline 
%Nd$_3$TlO   & ?  & ? & No & \textbf{No} & No  \\
%\hline 
%Pr$_3$InO   & No & Yes  & No & Yes & materialsproject.org  \\
%\hline
%Pr$_3$TlO   & ?  & ? & No & \textbf{No} &  No \\
%\hline
%\hline
%Ca$_3$BiN  & Yes & ? & No & Yes &  Ref.~\onlinecite{Sun2010}(3DTI under strain)\\
%\hline
%Ba$_3$BiN   & Yes & ? & No & Yes &  Ref.~\onlinecite{Sun2010}(3DTI under strain) \\
%\hline 
%Sr$_3$BiN   & Yes & ? & No & Yes &  Ref.~\onlinecite{Sun2010}(3DTI under strain) \\
%\hline
%\hline
%Cu$_3$PdN   & Yes  & No & Yes &  Refs.~\onlinecite{Yu_prl2015,Kim_prl2015}\\
%\hline
\end{tabular}
\label{tab:lanthanides}
\end{table}

\begin{figure*}[ht!]
\centering
\includegraphics[width=0.8\linewidth,clip=true]{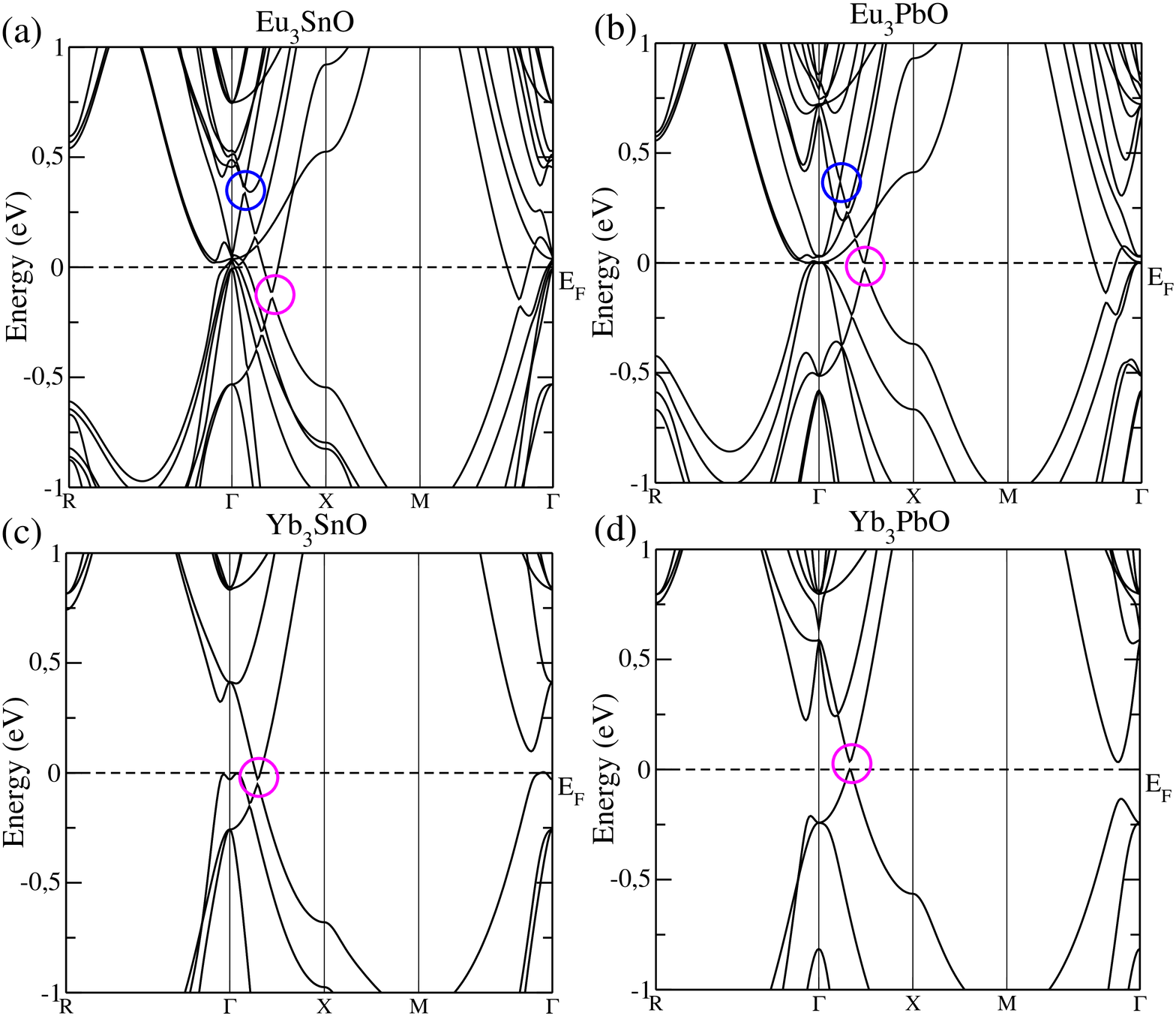}
\caption{Calculated bandstructures of cubic lanthanide antiperovskites (a) Eu$_3$SnO, (b) Eu$_3$PbO, (c) Yb$_3$SnO, and (d) Yb$_3$PbO. 
SOC is included in the calculations. The horizontal dashed line marks the position of the Fermi level. Circles highlight possible 
massive Dirac/Weyl fermion features along the $\Gamma$-X direction.}
\label{fig1}
\end{figure*} 

Based on bulk electronic structure, we conclude that  
Eu$_3$BO and Yb$_3$BO (B=Sn, Pb) are 3D DMs with a finite Dirac gap. The 
calculated bandstructures of these materials are presented in Fig.~\ref{fig1}.
Here %in both Table~\ref{table:alkali} and \ref{table:lanthanides}, 
we refer to 3D DM as materials having nodes (in general, gapped) in their 3D Brillouin zone, regardless of the nature of the 
nodes and their degeneracy. However, there is an important distinction 
between non magnetic (Yb$_3$BO) and magnetic (Eu$_3$BO) systems. Since magnetic ground state breaks time-reversal symmetry, 
we expect the degeneracy of the Dirac nodes to be lifted in magnetic lanthanide antiperovskites, giving rise to gapped Weyl nodes. 
Below we will discuss in more detail the bandstructures for these two cases.

Yb$_3$SnO and Yb$_3$PbO do not show any sizable magnetic moment, which is expected since Yb has a fully filled $f$-shell. These materials are characterized by 
a single node located along the $\Gamma$-X symmetry direction (X is the center of the face in the perovskite structure). Due to cubic symmetry, there are in total six nodes along the six equivalent $\Gamma$-X 
 paths parallel to $\pm\mathbf{a}$, $\pm\mathbf{b}$ and $\pm\mathbf{c}$, where $\mathbf{a}$, $\mathbf{b}$, $\mathbf{c}$ are reciprocal lattice vectors. 
 The degeneracy of the bands forming the nodes is $g=2$. Hence we identify the nodes in Yb$_3$BO as Dirac nodes. 
The energy gap at the node is $\Delta\approx 24$~meV for Yb$_3$SnO and $\Delta\approx 40$~meV for Yb$_3$PbO. 
The case of Yb$_3$PbO is particularly interesting as it has an isolated Dirac node at the Fermi level. In Yb$_3$SnO, the node is located slightly below the Fermi level.
 %(in the energy window of $\approx{100}$~meV there are no other electronic states present except the Dirac states). 

In our spin-polarized DFT calculations for magnetic lanthanide perovskites, we considered a ferromagnetic ground state, 
e.g. we assumed that magnetic moments of the lanthanide atoms, e.g. Eu in Eu$_3$BO (B=Sn, Pb), 
are parallel to each other. The calculated magnetic moments are equal for all magnetic atoms in the unit 
cell and are close to the magnetic moments of the cations, e.g. $\approx{7}$ $\mu_\mathrm{B}$ for Eu$^{2+}$.
 This is the simplest and the most feasible choice for our computational approach and it is partly justified by 
experimental measurments~\cite{Abdolazimi_thesis}. Such calculations could also represent a paramagnetic phase in external 
magnetic field, where all spins align parallel to each other.

It should be mentioned that manetic coupling in realistic materials is likely to be more complicated.      
In the cubic antiperovskite structure, the lanthanide cations are located  at the face-centrered 
positions, forming the cental octahedron [see Fig.~\ref{fig0}(a)]. Hence, the lattice is formed by triangular net of the magnetic moments. 
Such structure is prone to magnetic frustation and can lead to a complicated magnetic state with many possible phases. 
This has been documented in other materials with similar structure, for example in Mn-based antiperovskite nitrides, 
Mn$_3$XN (X= Ga, Zn, Cu, Ge, Sn), which display multiple magnetic interactions below the temperature of the paramegnetic 
transition.~\cite{Sun_jphys2012}
%It is likely that lanthanide antiperovskites behave similarly, although much less is known about magnetism in this materials.} 
 
An experimental study of magnetism and transport in Eu-based antiperovskites has been reported in Ref.~\cite{Abdolazimi_thesis}.
Interestingly, the magnetization measurements in Eu$_3$BO (B=Sn, Pb) did not confirm the anticipated magnetic 
frustration (the frustration factor extracted from the temperature dependence of the magnetization was found to be close to unity, meaning no frustration). 
At the same time, the materials become antiferromagnetic at low temperatures, with the Neel temperature of 33~K in Eu$_3$SnO and 43~K for Eu$_3$PbO. 
To reconcile antiferromagnetism with the apparent absence of frustration, a model of magnetic periodic sublattices was suggested. 
In this model, the crystal consists of parallel sublattices (planes), inside which the magnetic moments of lanthanide ions interact 
ferromagnetically, while between the planes the moments interact antiferromagnetically. Magnetic sublattices appear as steps in the 
field-dependent magnetization curves which has been confirmed in the experiment. The resulting phase diagram for Eu$_3$BO (B=Sn, Pb) 
reveals several magnetic ordered states with temperature- and field-dependent phase transitions between the states.
%In order to model this complicated magnetic order, more sophisticated electronic structure calculations are required and it is 
%outside the scope of the present work.}

Based on the assumption of a ferromagntic ground state, we find that Eu$_3$SnO and Eu$_3$PbO have a total magnetic moment of $\approx 21$~$\mu_\mathrm{B}$ and 
a magnetic moment on Eu site of $\approx 6.9$~$\mu_\mathrm{B}$. The direction of the magnetic moment is along $[001]$. Based on a simple model 
of a $4\times{4}$ Dirac Hamiltonian in external magnetic field, the Dirac node is split into two Weyl nodes, separated in momentum space by 
a vector proportional to the magnetic field vector~\cite{footnote}  Assuming that the magnetization acts as an effective magnetic field, we expect the shifted Weyl nodes to appear along the $\Gamma$-$X$ 
direction parallel to $+\mathbf{c}$, which we refer to as $\Gamma$-X$_c$. The path used for bandstructure calculations presented in Fig.~\ref{fig1} includes 
$\Gamma$-X$_c$ [see Fig.~\ref{fig0}(b)].

\begin{figure}[ht!]
\centering
\includegraphics[width=0.8\linewidth,clip=true]{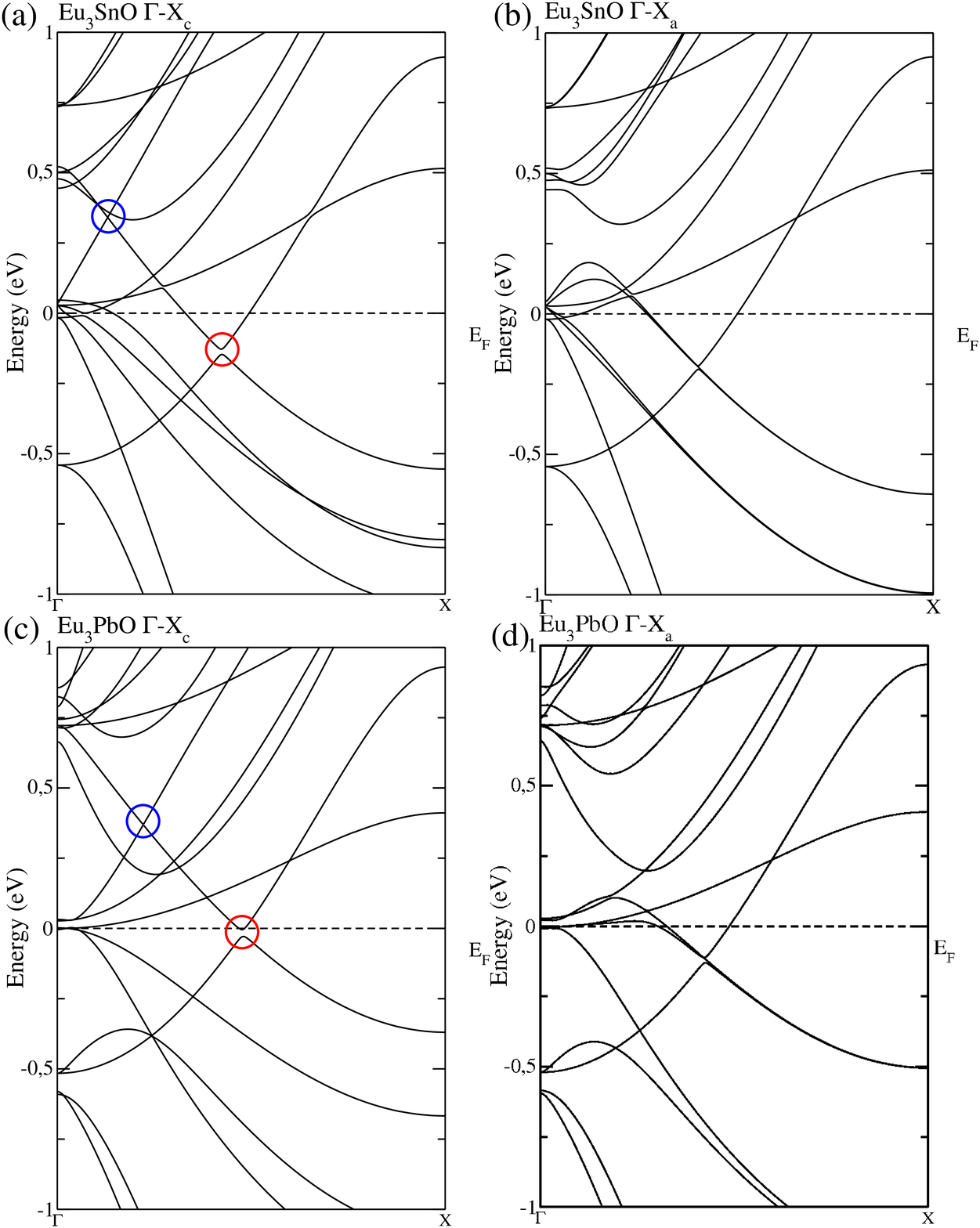}
\caption{Calculated bandstructures of magnetic lanthanide antiperovskites (a,b) Eu$_3$SnO and (c,d) Eu$_3$PbO 
along two non-equivalent paths $\Gamma$-X$_c$ (left) and $\Gamma$-X$_a$ (right). Possible Weyl nodes shifted along the 
$\Gamma$-X$_c$ direction are marked by circles in panels (a) and (c).}
\label{fig1a}
\end{figure}

\begin{figure*}[ht!]
\centering
\includegraphics[width=0.8\linewidth,clip=true]{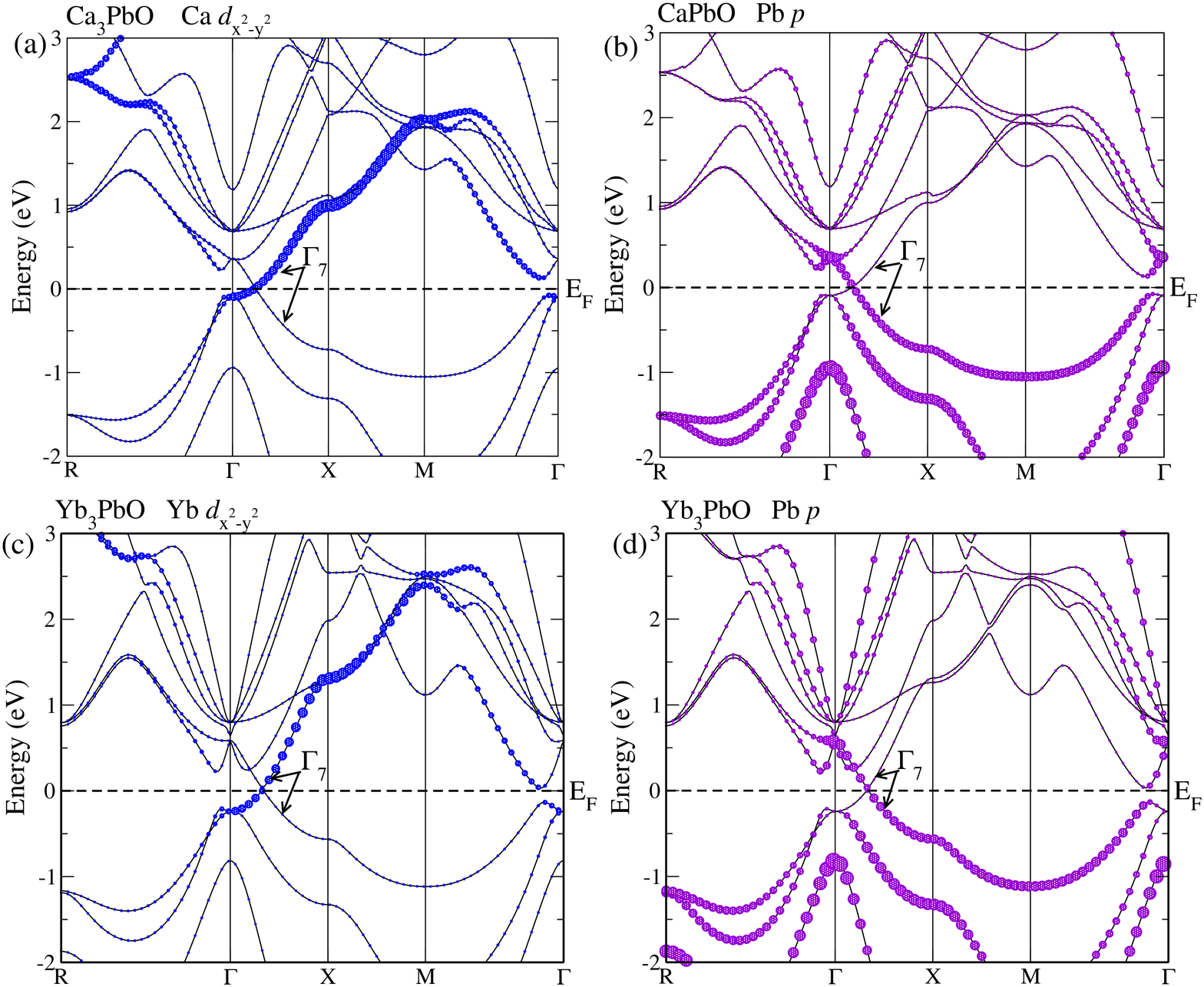}
\caption{Band character of cubic alkali antiperovskite (a,b) Ca$_3$PbO and lanthanide antiperovskite (c,d) Yb$_3$PbO. 
Filled circles show relative contributions of Ca and Yb $d$-states (a,c) and Pb $p$ states (b,d). Only $d_{x^2-y^2}$ states are shown for 
Ca and Yb. We have verified that the contribution of other $d$ orbitals with different symmetry is 
negligible at low energies. 
 In panels (c) and (d), the highly localized Yb 4f states are located below -2eV. The horizontal dashed line marks the position of the Fermi level. 
 Irreducible representations are shown for valence and conduction bands  along the $\Gamma$-X direction.}
\label{fig2}
\end{figure*}

One can see in Fig.~\ref{fig1}(a,b) that Eu$_3$BO indeed displays nodes along the $\Gamma-X_c$ directions. 
There is a gapped node, which is similar in shape and location to Yb$_3$BO, located at the Fermi level in Eu$_3$PbO and at $\approx{100}$~meV below the Fermi level 
in Eu$_3$SnO [marked with magenta circle in Fig.~\ref{fig1}(a,b)]. The energy gap at the node is $\Delta\approx 25$~meV. 
Apart from this node, there are many other avoided band crossings near the Fermi level. One possible node at $\approx{300}$~meV above the Fermi level 
is marked with blue circles in Fig.~\ref{fig1}(a,b). We find a smaller energy gap at this node, $\Delta\approx 1$~meV for Eu$_3$SnO and $\Delta\approx 10$~meV for Eu$_3$PbO. 
 To further clarify the position of the nodes, we plot in Fig.~\ref{fig1a}
the bandstructures of Eu$_3$BO along two non-equivalent paths, $\Gamma$-X$_c$ and $\Gamma$-X$_a$ (parallel to $+\mathbf{a}$). The node at positive energy (blue circles)  
is clearly absent in the bandstructure plotted along $\Gamma$-X$_a$. 
We verified that the band degeneracy around the gapped nodes is $g=1$. 
The number of nodes and the degeneracy indicate that the nodes in Eu$_3$BO could be identified as Weyl nodes.

In order to precisely identify the Weyl nodes, further bulk and surface calculations are 
required. This is outside the scope of the present work since our goal is to identify potential 3D DM candidates. 
It is worth pointing out that the expectations based on the $4\times{4}$ Dirac Hamiltonian 
do not necessarily hold in realistic materials. Villanova and Park~\cite{Villanova_prb2018} showed, using an \textit{ab-initio}-based tight-binding 
model, that in Ni$_3$Bi Dirac semimetal the Dirac node is split into four Weyl nodes when magnetic field is applied. 
Moreover, a finite magnetization in intrinsically magnetic materials is not equivalent to external magnetic field.

Next, we comment on the possible origin of bulk Dirac states in lanthanide antiperovskite oxides. For this we compare the character of the 
low energy states for the typical alkali antiperovskite oxide Ca$_3$PbO 
 and Yb$_3$PbO (see Fig.~\ref{fig2}). 
It was noted in Refs.~\cite{Kariyado_jpsj2011, Kariyado_jpsj2012} that the low energy Dirac states in Ca$_3$PbO are formed mainly 
 by Ca $3d$ and Pb $6p$ orbitals. Among Ca $3d$, the $3d_{x^2-y^2}$ orbitals give the major contribution to the Dirac states.
 We have verified this picture [see the electronic structure of Ca$_3$PbO with band character in Fig.~\ref{fig2}(a,b)]. 
 
 Based on the insight from DFT electronic structure, Kariyado and Ogata~\cite{Kariyado_jpsj2011, Kariyado_jpsj2012} constructed a 
 tight-binding Hamiltonian with a basis formed by Ca $d_{x^2-y^2}$ and Pb $p$ states, which the authors proved to be a Dirac Hamiltonian yielding 
 3D gapless Dirac states along the $\Gamma$-$X$ line.  
It was suggested that a small gap at the Dirac node (few tens of meV's) is due to SOC and 
 the coupling to other states away from the Fermi level i.e. Pb $5d$ and other Ca $3d$ states~\cite{Kariyado_jpsj2012}. 
 We have verified in our calculations that in the absence of SOC, the Dirac states are gapless. %Therefore it is likely that SOC gives the major contribution to the gap. 
Further analysis of the DFT bands for Ca$_3$PbO with SOC reveals that the two bands forming the node along the $\Gamma$-$X$ direction belong to the same 
 irreducible representation, $\Gamma_7$ of the $C_{4v}$ double point group [see Fig.~\ref{fig2}(a,b)], which further justifies the avoided crossing between these bands 
 (irreducible representation have been computed with Wien2k~\cite{Wien2k} and independently verified with QUANTUM ESPRESSO~\cite{qe2009,qe2017}).  
 
 SOC is also responsible for the band inversion at the center of the Brillouin zone which was found in Ca$_3$PbO~\cite{Kariyado_jpsj2011, Kariyado_jpsj2012, Hsieh_prb2014}.  
 Since Ca $d$ and Pb $p$ states, which give rise to low-energy electronic states, have different parities,  
 there exist two possible band orderings at the $\Gamma$ point.
 In the normal ordering, the $d$ orbitals of Ca (A position in A$_3$BO)  
lie above the Pb $p$ orbitals (B position in A$_3$BO). However, in the presence of SOC the order is reversed, namely 
 the top of the $p$ bands lies above the bottom of the $d$ bands.  
 This band inversion signals a possible topological phase which was detected by Klintenberg \textit{et al.}~\cite{Klintenberg2014}. However, it leaves the materials 
 a trivial insulator in the Z$_2$ classification of TIs because the product of parity eigenvalues remains unchanged due to the four-fold degeneracy of valence and 
 conduction band extrema at $\Gamma$
 ~\cite{Sun2010,Kariyado_jpsj2012}. 
 
 Nevertheless, the inverted band ordering is an important ingredient in the low-energy effective Hamiltonian of Ref.~\cite{Kariyado_jpsj2012} which 
 produces 3D Dirac states away from $\Gamma$. It was subsequently demonstrated by Hsieh \textit{et al.}~\cite{Hsieh_prb2014} that this band inversion 
 leads to a TCI phase described by a nonzero mirror Chern number. 
 It was also shown that the bulk Dirac node is a direct consequence of the band inversion at $\Gamma$ i.e. the band crossing on the $\Gamma$-$X$ is absent in the trivial phase 
 characterized by normal band ordering~\cite{Hsieh_prb2014}. 
  This 
 is different from the band inversion at the $R$ point in Cu$_3$PdN, which involves $p$ and $d$ states of the metallic anion (Pd $5p$ states are lower
than Pd $4d$ states). When SOC is included, Cu$_3$PdN exhibits Dirac nodes along $R$-$X$ and $R$-$M$ directions, where the node along $R$-$M$ 
remains gapless~\cite{Yu_prl2015}. Similarly, the band inversion at $X$ in Cu$_3$ZnN leads to nodes along $R$-$X$ and $R$-$M$~\cite{Kim_prl2015}. 
 
 We find that the character of low energy states in Yb$_3$PbO is similar to Ca$_3$PbO [see Fig.~\ref{fig2}(c,d)]. In this case, the Dirac states are formed mainly by 
 Yb $4d_{x^2-y^2}$ and Pb $6p$ orbitals. The valence and conduction bands along $\Gamma$-$X$ belong to the same irreducible representation, $\Gamma_7$ of the $C_{4v}$ group.  
 These are strong indications that the origin of the Dirac states in this material is similar to 
 Ca$_3$PbO.  Hence, we conclude that Yb$_3$PbO is characterized by massive 3D Dirac fermions at finite $k$ and is also a potential TCI. 
 Further theoretical and experimental work will be useful to describe bulk Dirac states and possible topological surface states due to the TCI phase in this material 
 and other lanthanide antiperovskites with similar properties.  
 
\begin{figure}[ht!]
\centering
\includegraphics[width=0.8\linewidth,clip=true]{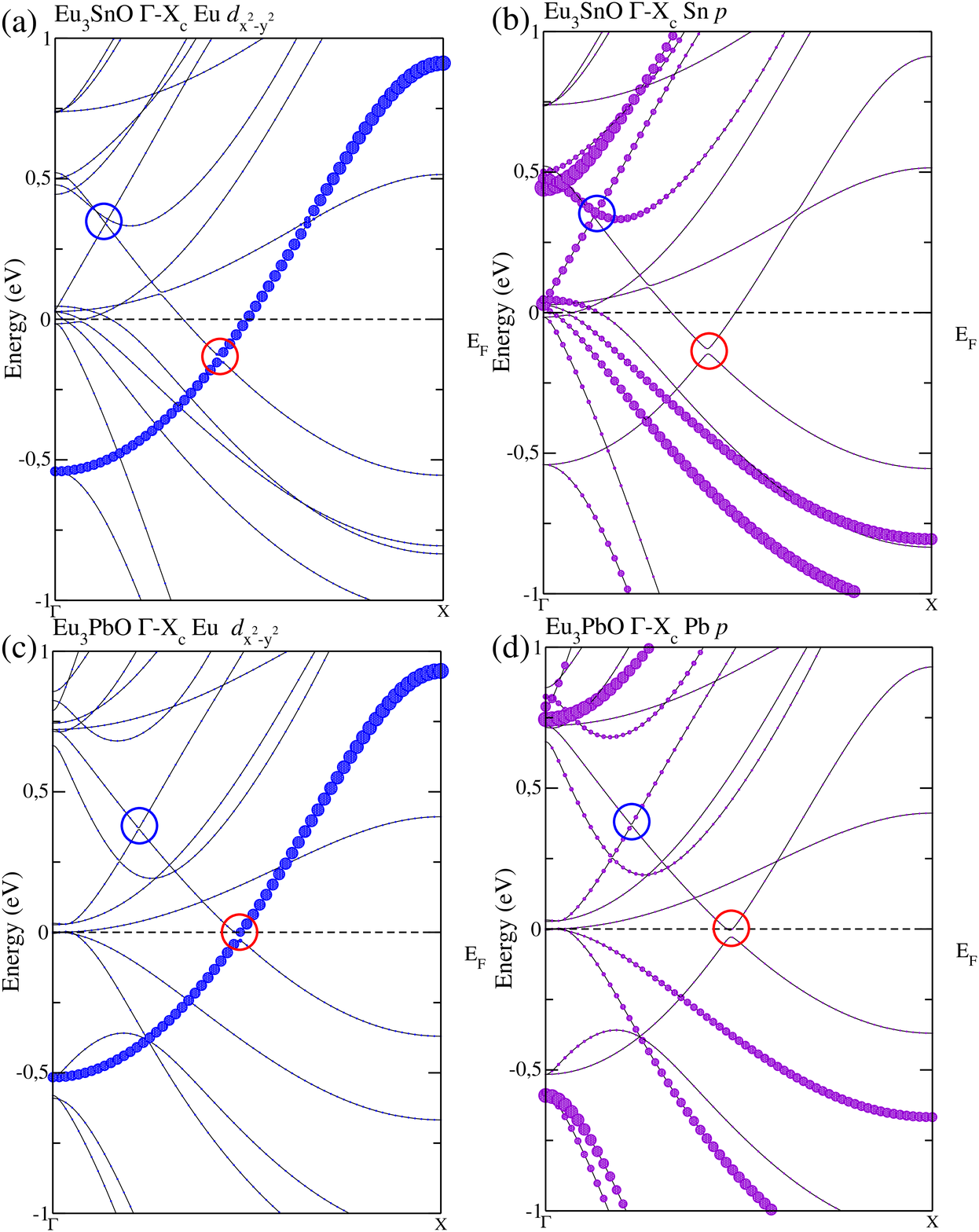}
\caption{Band character of magnetic antiperovskite (a,b) Eu$_3$SnO and (c,d) Eu$_3$PbO. 
Filled circles show relative contributions Eu $d$-states (a,c) and Sn/Pb $p$ states (b,d). Bands are plotted along the $\Gamma$-X$_c$ path.}
\label{fig3}
\end{figure}

Since the above discussion concerned non-magnetic antiperovskites, we also checked the character of the bands for magnetic 
Eu$_3$BO (see Fig.~\ref{fig3}). The features in the bandstructures that we identified as potential Weyl nodes do not have exactly the same band character as 
the Dirac node in Yb$_3$PbO. However, they do have contributions from Eu $d$ and Sn/Pb $p$ states.

Finally, we also considered hypothetical structures A$_3$BO,  A=Gd, Sm and B=Sn, Pb, for which no experimental structural data is available to date. 
For these compounds, we have obtained DFT optimized crystal structures (see Table~\ref{tab:lattice}), however their dynamic structural stability needs to be investigated further. 
Although our calculations showed interesting features, including band crossings near the Fermi level, we were not able to clearly identify 
3D Dirac states in these compounds.

\section{Conclusion}\label{concl}

In conclusion, we performed an \textit{ab initio} search for Dirac materials among some antiperovskite oxides. 
Our data set includes both the Ca$_3$PbO family as a test subset and the less studied lanthanide antiperovskites 
containing $f$-electron elements, some of which have not been synthesized to date. 
For the Ca$_3$PbO family (A$_3$BO, A=Ca, Ba, Sr, B=Sn, Pb), our electronic structure calculations show 
 gapped 3D Dirac states, which is consistent with previous theoretical work. 
Among the $f$-electron antiperovskites, we identify Eu$_3$BO and Yb$_3$BO (B=Sn, Pb) as 3D DMs, which display massive Dirac nodes along the $\Gamma$-$X$ 
in the 3D Brillouin zone. 

In Yb$_3$BO (B=Sn, Pb) the nodes are energetically close to the Fermi level and the energy gap at the nodes is few tens of meV's. 
 Yb$_3$PbO is particularly promising as it has an isolated Dirac node at the Fermi level. 
Eu$_3$BO (B=Sn, Pb) possess a finite magnetic moment due to unfilled $f$-electron shell. 
We find that in these magnetic lanthanide antiperovskites the degeneracy of the Dirac nodes can be lifted, leading to 
appearance of Weyl nodes in the 3D Brillouin zone. 
 As demonstrated previously for Ca$_3$PbO family, the existence of bulk Dirac states at finite \textit{k} is a direct consequence of the band inversion 
 at the center of the 3D Brillouin zone, which gives rise to a TCI phase.   
 Hence, the 3D DMs found among the $f$-electron antiperovskites may be potential TCIs.
 
\section{Acknowledgments}
We are grateful to P. Hofmann for useful discussion. This work was supported by the VILLUM FONDEN 
via the Centre of Excellence for Dirac Materials (Grant No. 11744), the European Research Council 
under the European Union’s Seventh Framework Program (FP/2207-2013)/ERC Grant Agreement No.~DM-321031, the 
Swedish Research Council Grant No.~638-2013-9243, and the Knut and Alice Wallenberg Foundation. The authors 
acknowledge computational resources from the Swedish National Infrastructure for Computing (SNIC) at the High 
Performance Computing Center North (HPC2N)
\bibliography{Driven_Dirac}

\begin{thebibliography}{42}
\bibitem [{foo()}]{footnote}%
  \BibitemOpen
  \href@noop {} {}\bibinfo {howpublished} {Consider the Dirac Hamiltonian
  $H_\mathrm{D}=\tau_z\otimes{v}_\mathrm{F}\mathbf{k}\cdot\mathbf{\sigma}$,
  where $v_\mathrm{F}$ is the Fermi velocity, $\mathbf{\sigma}$ and
  $\mathbf{\tau}$ are Pauli matrices. Interaction with an effective magnetic
  field $\mathbf{b}$ can be written as
  $H_\mathrm{b}=-\tau_0\otimes\mathbf{\sigma}\cdot\mathbf{b}$. Then the
  eigenstates of the Hamiltonian $H=H_\mathrm{D}+H_\mathrm{b}$ are given by
  $E_{\pm}^{\xi}=\pm{v}_\mathrm{F}|(\mathbf{k}-\xi\mathbf{k}_0)|$, where
  $\xi=\pm{1}$ and $\mathbf{k}_0=\mathbf{b}/{v}_\mathrm{F}$. This gives two
  Weyl nodes with chiralities $\xi$, separated by vector
  $2\mathbf{k}_0$.}\BibitemShut {Stop}%
\end{thebibliography}

\end{document}